\documentstyle[a4,epsf]{article}
\title{\bf Spontaneous CP violation model \\ with flavor symmetry \\
 in large extra dimensions}
\author{{\bf Yutaka Sakamura}\footnote{Email: sakamura@th.phys.titech.ac.jp}\\
{\footnotesize\it Department of Physics, Tokyo Institute of Technology} \\
{\footnotesize\it Oh-okayama, Meguro, Tokyo 152-8551, Japan}}
\date{}
\newcommand{\ep}{\epsilon}
\newcommand{\tilep}{\tilde{\epsilon}}
\newcommand{\mgut}{M_{\rm\footnotesize GUT}}
\newcommand{\bphi}{\mbox{\boldmath $\Phi$}}
\begin{document}
\maketitle
\begin{abstract}
We construct the minimal SUSY model that causes spontaneous CP violation 
with an abelian flavor symmetry in the context of the large extra dimensions 
and show that various phenomenological problems can be solved by introducing
only scales below the low fundamental scale.
We also realize the realistic size of the CP violation and the small masses
of the neutrinos.
The strong CP problem can be solved by the axion scenario and the axion can
be made invisible by introducing an additional large extra dimension.
\end{abstract}

\section{Introduction}
There is a large hierarchy among the fermion masses that have been observed.
This hierarchy cannot be explained within the standard model and thus
tackling this problem gives crucial clues to high energy physics beyond
the standard model.
One of the solution to this problem is introducing a flavor symmetry, 
which acts on fermions in a flavor dependent way \cite{fro-niel}.

Another problem that has not been solved is the origin of the CP violation.
CP symmetry is a very good symmetry, but it has been observed 
that CP is violated in the neutral K meson system by a small amount.
This smallness of the CP violation is also a mystery in particle physics.
One of the convincing solution to this problem is the spontaneous CP violation
(SCPV) \cite{lee}.
This idea is very attractive since
it can control the small amount of the observed CP violation quite naturally.

Recently the impact of the existence of the large extra dimensions,
which is indicated by the string theory \cite{anton}, are discussed 
in many papers.
The main impact of this possibility is that the fundamental scale of the
theory $M_{\ast}$ can be lowered from the Planck scale 
($\sim 10^{19}$~GeV) to a much lower scale \cite{lykken,arkani}.
This argument is assumed that only gravitons can propagate into the large 
extra dimensions and the standard model particles are confined to 
a four-dimensional hypersurface, such as a D3-brane \cite{rubakov}.
In the case that the gauge bosons of the standard model 
(and some of the matter particles) also propagate into the extra 
dimensions, we can lower the grand unification scale $\mgut$ to the TeV
region due to the power-law running of the gauge couplings \cite{dienes}.

Another implication of the large extra dimensions is the volume factor 
suppression of the couplings between the bulk fields and boundary fields,
which is used to explain the smallness of the neutrino masses 
in Ref.\cite{dienes2}.
This suppression was also used to explain the hierarchy among
the masses of quarks and leptons \cite{yoshioka,sakamura}\footnote{
Recently, however, it is shown that this possibility seems to be disfavored 
in the context of the TeV-strings because the Kaluza-Klein modes of the gluons
generate dangerous flavor and CP-violating interactions.}.

In this paper, we discuss the minimal model in which the fermion mass 
hierarchy is realized by an abelian flavor symmetry and the smallness of the CP
violation is controlled by the SCPV scenario in the context of the large extra
dimensions.
Then we will show that various phenomenological problems can be solved by
introducing only scales below the low fundamental scale (1000~TeV in our 
scenario).

The paper is organized as follows.
In the next section, we will introduce our model and realize the realistic
hierarchy among fermion masses.
In Section~\ref{scpv}, we will estimate CP violation parameters and show
that they are consistent with the experimental values, and
other phenomenological implications are discussed in Section~\ref{opi}.
Section~\ref{concl} is devoted to conclusions.

\section{Model}
\subsection{Flavor symmetry and Yukawa hierarchy}
We will introduce an abelian global symmetry $U(1)_{A}$ as the flavor symmetry.
Here we will deal with the 4-Higgs-doublet (4HD) model because it is 
the minimal model that causes SCPV as is explained later.
Thus some symmetry is required to suppress the dangerous flavor changing 
neutral current (FCNC) and we will introduce the $Z_{2}$ symmetry, which 
distinguishes the ``standard'' Higgs doublets that mainly give masses to 
fermions from the ``extra'' Higgs doublets that are prevented from having 
vacuum expectation values (VEVs).
Since the fundamental scale $M_{\ast}$ is around 1000~TeV in our scenario,
the supersymmetry (SUSY) is assumed to avoid the naturalness problem.
R-parity is assumed here.
We will consider the case that there is a large extra dimension, which is  
compactified by the radius $R$, besides 
the usual four-dimensional space-time and some fields feel the fifth dimension
while the others are confined to a four-dimensional boundary.
The field contents of our model and their charges of the $U(1)_{A}$
and $Z_{2}$ parity are listed in Table~\ref{4dfld} and Table~\ref{5dfld}.
Here $Q_{i}$, $\bar{U}_{i}$, $\bar{D}_{i}$, $L_{i}$ and $\bar{E}_{i}$
represent the superfields of the $i$-th generation of the left-handed quark 
doublet, 
right-handed up-type quark singlet, right-handed down-type quark singlet, 
left-handed lepton doublet and right-handed charged lepton singlet,
respectively.
$A_{\mu}^{a}$ ($a=1,2,3$) are the gauge superfields and 
$H_{i}$ ($i=1,2,3,4$) are the superfields of the Higgs doublets.
They are all the description from the viewpoint of the four-dimensional 
theory, and the superfields listed in Table~\ref{5dfld} should be interpreted 
as the Kaluza-Klein zero modes for the extra dimension.
$\bphi$, $\bphi'$ and $\bphi''$ denote the five-dimensional gauge-singlet 
scalar fields and their superpartners.

\begin{table}[t]
\begin{center}
\begin{tabular}{|c|c||c|c||c|c||c|c||c|c|} \hline
 $Q_{1}$ & 2 (+) & $\bar{\bar{U}_{1}}$ & 2 (+) & $\bar{E}_{1}$ & 2 (+) & 
 $H_{1}$ & $-1$ (+) & $H_{3}$ & $-1$ ($-$) \\ \hline
 $Q_{2}$ & 1 (+) & $\bar{\bar{U}_{2}}$ & 1 (+) & $\bar{E}_{2}$ & 1 (+) & 
 $H_{2}$ & 0 (+) & $H_{4}$ & 0 ($-$) \\ \hline
 $Q_{3}$ & 0 (+) & $\bar{\bar{U}_{3}}$ & 0 (+) & $\bar{E}_{3}$ & 0 (+) & 
 & & & \\ 
\hline
\end{tabular}
\caption{The fields confined to the four-dimensional boundary and 
 their $U(1)_{A}$ charges. The numbers in the parentheses are $Z_{2}$ parity.} 
 \label{4dfld}
\end{center}
\begin{center}
\begin{tabular}{|c|c||c|c||c|c||c|c|} \hline
 $\bar{\bar{D}_{1}}$ & 1 (+) & $L_{1}$ & 1 (+) & $A_{\mu}^{1}$ & 0 (+) & 
 $\bphi$ & $-1$ (+) \\ \hline
 $\bar{\bar{D}_{2}}$ & 1 (+) & $L_{2}$ & 1 (+) & $A_{\mu}^{2}$ & 0 (+) & 
 $\bphi'$ & 3 (+) \\ \hline
 $\bar{\bar{D}_{3}}$ & 1 (+) & $L_{3}$ & 1 (+) & $A_{\mu}^{3}$ & 0 (+) & 
 $\bphi''$ & 0 ($-$) \\ \hline
\end{tabular}
\caption{The fields that can propagate into the extra dimension and 
 their $U(1)_{A}$ charges. The numbers in the parentheses are $Z_{2}$ parity.} 
 \label{5dfld}
\end{center}
\end{table}

Here we will give a brief explanation of the volume factor suppression.
Let us denote $\Psi(x,y)$ as a five-dimensional bulk field, where $y$ 
represents the coordinate of the extra dimension.
If we Fourier expand
\begin{equation}
 \Psi(x,y)=\sum_{m=0}^{\infty}\frac{1}{\sqrt{2\pi R}}\psi_{m}(x)e^{i(m/R)y},
\end{equation}
then  we can regard $\psi_{m}(x)$ as a four-dimensional field corresponding
to the $m$-th Kaluza-Klein mode.

On the other hand, the boundary fields are localized at the four-dimensional
wall whose thickness is of order $M_{\ast}^{-1}$,\footnote{
We consider the hard brane case, in which the brane tension is of order 
$M_{\ast}^{4}$ \cite{bando}.} so a coupling involving
at least one boundary field is suppressed by a factor of 
$(1/\sqrt{2\pi M_{\ast}R})^{k}$, where $k$ is a number of bulk fields
included in the coupling \cite{dienes2}.

With the above charge assignment, we have the Yukawa couplings with 
the following structure after the scalar fields $\bphi$ and $\bphi''$ obtain 
the VEVs,

\begin{eqnarray*}
 W_{\rm yukawa}&=&-h^{d}_{ij}Q_{i}\bar{D}_{j}H_{1}
 +h^{u}_{ij}Q_{i}\bar{U}_{j}H_{2}-h^{e}_{ij}L_{i}\bar{E}_{j}H_{1} \\ \nonumber
 &&-h^{'d}_{ij}Q_{i}\bar{D}_{j}H_{3}
 +h^{'u}_{ij}Q_{i}\bar{U}_{j}H_{4}-h^{'e}_{ij}L_{i}\bar{E}_{j}H_{3},
\end{eqnarray*}
\begin{displaymath}
 h^{u}_{ij}\simeq\left(\begin{array}{ccc}\ep^{4} & \ep^{3} & \ep^{2} \\
 \ep^{3} & \ep^{2} & \ep \\ \ep^{2} & \ep & 1 \end{array}\right),\;\;
 h^{d}_{ij}\simeq\frac{1}{\sqrt{2\pi M_{\ast}R}}
 \left(\begin{array}{ccc}\ep^{2} & \ep^{2} & \ep^{2} \\
 \ep & \ep & \ep \\ 1 & 1 & 1 \end{array}\right),
\end{displaymath}
\begin{equation}
 h^{e}_{ij}\simeq\frac{1}{\sqrt{2\pi M_{\ast}R}}
 \left(\begin{array}{ccc}\ep^{2} & \ep & 1 \\
 \ep^{2} & \ep & 1 \\ \ep^{2} & \ep & 1 \end{array}\right),\;\;
 h^{'u}\simeq\tilep h^{u},\;\; h^{'d}\simeq\tilep h^{d},\;\; 
 h^{'e}\simeq\tilep h^{e}, \label{yukawa}
\end{equation}
where
\begin{equation}
 \ep\equiv\left|\frac{\langle\bphi\rangle}{M_{\ast}^{3/2}}\right|
 =\left|\frac{v}{M_{\ast}}\right|^{3/2},\;\;
 \tilep\equiv\left|\frac{\langle\bphi''\rangle}{M_{\ast}^{3/2}}\right|
 =\left|\frac{v''}{M_{\ast}}\right|^{3/2}.
\end{equation}

Here $\langle\bphi\rangle\equiv v^{3/2}$ and $\langle\bphi''\rangle\equiv 
v^{\prime\prime 3/2}$.
Note that the bulk fields $\bphi$ and $\bphi''$ have the mass dimension of 3/2.

Eq.(\ref{yukawa}) suggests that $\sqrt{2\pi M_{\ast}R}\simeq m_{t}/m_{b}\simeq 40$, so that
$M_{\ast}\simeq 250R^{-1}$.
If $(v/M_{\ast})^{3/2}\simeq 1/15$, i.e., $v\simeq M_{\ast}/6$, 
the realistic hierarchy between fermion masses can be realized
\footnote{It can easily be shown that the runnings of Yukawa couplings
between $\mgut$ and $R^{-1}$ do not destroy this hierarchy.}.

The value of $R^{-1}$ is constrained from Ref.\cite{masip-pomarol,del-pom-qrs}
to be greater than 2~TeV, so we will set $R^{-1}=4$~TeV throughout this paper
\footnote{The relation between the fundamental scale $M_{\ast}$ and the Planck
scale $M_{p}$ in the case of the $n$ extra dimensions is
$M_{p}^{2}=(2\pi)^{n}M_{\ast}^{2+n}R_{1}R_{2}\cdots R_{n}$, where $R_{i}$ is
the radius of the $i$-th extra dimension.
We have assumed the existence of more extra dimensions that are irrelevant to
our discussion to satisfy this relation.}.
Therefore $M_{\ast}\simeq 1000$~TeV, $v\simeq 160$~TeV and $\ep\simeq 1/15$.

\subsection{Dangerous FCNC}
We can expand the bulk field $\bphi$ around its VEV in terms of 
four-dimensional fields $\phi_{n}$ as follows.

\begin{equation}
 \bphi(x,y)=v^{3/2}+\frac{1}{\sqrt{2\pi R}}\sum_{n=0}^{\infty}\phi_{n}(x)
 e^{i(n/R)y},
\end{equation}
where $x$ and $y$ represent the four-dimensional coordinates and 
the coordinate of the extra dimension respectively.

These fluctuation fields $\phi_{n}$ are thought to have 
masses of order $v$ if $n$ is not very large.
Since these are much lighter than the counterpart of the usual 
Froggatt-Nielsen mechanism \cite{fro-niel}, tree-level processes exchanging
these light $\phi_{n}$ seem to cause the disastrously large FCNC 
at first sight.
However, their couplings to the light quarks and leptons are largely suppressed
by the large power of the volume factor $1/\sqrt{2\pi M_{\ast}R}\simeq 1/40$,
so there are no dangerous FCNC processes arising from $\phi_{n}$-exchange.

There are another flavor changing interactions that might give rise to
too large FCNC.
In the case of an abelian flavor symmetry, flavor changing 4-fermi interactions
such as $(q_{1}\bar{d}_{2})(\bar{q}_{1}d_{2})$, where $q_{i}$ and $\bar{d}_{i}$
are quark fields, are allowed \cite{arkani2}.
In our case, however, taking into account the volume factor suppression,
this flavor changing interaction terms are

\begin{equation}
 {\cal L}_{FC}=\frac{a}{M_{\ast}^{2}(2\pi M_{\ast}R)}
 (q_{1}\bar{d}_{2})(\bar{q}_{1}d_{2})+\cdots,
\end{equation}
where $a$ is a dimensionless $O(1)$ constant.

Thus an effective cut-off $\Lambda$ becomes 
$\Lambda\sim M_{\ast}\sqrt{2\pi M_{\ast}R}\sim 4\times 10^{4}$~TeV 
and the FCNC arising from these interactions do not 
exceed the experimental bounds. (See Table~1 in Ref.\cite{arkani2}.)

\subsection{Grand Unification}
In the case that the gauge bosons of the standard model propagate into the 
large extra dimensions, the grand unification scale $\mgut$ is significantly 
lowered \cite{dienes}.
The new GUT scale is expected to be $20\sim 30R^{-1}$ from Ref.\cite{dienes}.
Unfortunately, the naive unification of the gauge couplings does not occur
in our model, but the cross points of the running of the gauge couplings
are not so far each other due to their power-law runnings.
Taking this fact into account, the grand unification of the gauge couplings
might be realized with the help of the threshold effect at $\mgut$.
In this paper, we will assume such a situation and use the term ``GUT'' 
in this sense.

Note that $\mgut\simeq 20\sim 30R^{-1}$ is the same order as $v$.
This suggests that $\bphi$ obtains the VEV at $\mgut$.
It seems quite natural for $\bphi$ to acquire the VEV at $\mgut$
because at the same scale some scalar fields obtain non-zero
VEVs and break the GUT gauge group into the standard model group: 
$SU(3)_{C}\times SU(2)_{L}\times U(1)_{Y}$.

\section{Spontaneous CP violation} \label{scpv}
\subsection{Higgs sector}
We will assume that the bulk scalar fields $\bphi$, $\bphi'$ and $\bphi''$
have complex VEVs at $\mgut$ and thus CP is violated spontaneously.
The effective $\mu$-terms are generated after the bulk scalars obtain 
the VEVs.
\begin{eqnarray}
 &&W_{\mu{\rm term}}= \nonumber \\
 &&\lambda_{12}v\left(\frac{v}{M_{\ast}}\right)^{2}
 \left(\frac{v'}{M_{\ast}}\right)^{3/2}H_{1}H_{2}
 +\lambda_{14}v\left(\frac{v}{M_{\ast}}\right)^{2}\left(\frac{v'}{M_{\ast}}
 \right)^{3/2}\left(\frac{v''}{M_{\ast}}\right)^{3/2}H_{1}H_{4} \nonumber \\
 &&+\lambda_{32}v\left(\frac{v}{M_{\ast}}\right)^{2}\left(\frac{v'}{M_{\ast}}
 \right)^{3/2}\left(\frac{v''}{M_{\ast}}\right)^{3/2}H_{3}H_{2}
 +\lambda_{34}v\left(\frac{v}{M_{\ast}}\right)^{2}\left(\frac{v'}{M_{\ast}}
 \right)^{3/2}H_{3}H_{4}, \nonumber \\
\end{eqnarray}
where $\lambda_{ij}$ are dimensionless $O(1)$ real couplings.

From now on, we will set $v$ to real by using the $U(1)_{A}$ symmetry and
assume that $|v|\simeq |v'|\simeq 40R^{-1}=160$~TeV.
In this case,
\begin{equation}
 \left|v\left(\frac{v}{M_{\ast}}\right)^{2}\left(\frac{v'}{M_{\ast}}\right)
 ^{3/2}\right|\simeq 0.07R^{-1}\simeq 300\;\,{\rm GeV},
\end{equation}
so the weak-scale $\mu$-parameters are generated.

As a result, the $\mu$-terms of the superpotential have the following 
structure.
\begin{equation}
 W_{\mu{\rm term}}=\mu_{12}e^{i\alpha}H_{1}H_{2}
 +\tilep\mu_{14}e^{i\beta}H_{1}H_{4}
 +\tilep\mu_{32}e^{i\beta}H_{3}H_{2}+\mu_{34}e^{i\alpha}H_{3}H_{4},
\end{equation}
where $\mu_{ij}$ are weak-scale order and real, and the phases $\alpha$ and 
$\beta$ are of order one.
The value of $\tilep\equiv|v''/M_{\ast}|^{3/2}$ is determined later.

\subsection{soft SUSY breaking parameters}
We will apply the Scherk-Schwarz mechanism \cite{scherk} to break 
the supersymmetry.
In this case the SUSY breaking scale is identified with the compactification
scale $R^{-1}$.
According to the charge assignment of the R-parity, the particles have
the following masses below the SUSY breaking scale $R^{-1}$ \cite{anton2}.

\begin{description}
 \item[(i) SM particles (except Higgs bosons) :] massless
 \item[(ii) Higgs bosons :] obtain masses through one-loop
 \item[(iii) gaugino, $\tilde{\bar{d}}$ and $\tilde{l}$ :] $O(R^{-1})$ 
  ($\sim 2$~TeV)
 \item[(iv) the other sfermions :] 
  $O(\sqrt{\alpha_{1}}R^{-1}\sim\sqrt{\alpha_{3}}R^{-1})$ (several hundred GeV)
\end{description}
where $\tilde{\bar{d}}$ and $\tilde{l}$ are the scalar components of
$\bar{D}$ and $L$ respectively.

The mass (squared) matrices of the sfermions satisfy the degeneracy and 
proportionality conditions, which are required to suppress the dangerous 
FCNC \cite{nir}.

In particular, the mass parameters in the Higgs sector are
\begin{equation}
 m_{i}^{2}\simeq (\mbox{a few hundred GeV})^{2},\;\;
 m_{ij}^{2}\simeq \mu_{ij}\alpha R^{-1}.
\end{equation}
Thus the Higgs potential $V$ has the following structure,
\begin{eqnarray}
 V&=&m_{1}^{2}H_{1}^{\dagger}H_{1}+m_{2}^{2}H_{2}^{\dagger}H_{2}
 +m_{3}^{2}H_{3}^{\dagger}H_{3}+m_{4}^{2}H_{4}^{\dagger}H_{4} \label{pottl}
\\ \nonumber
 &&+(m_{12}^{2}H_{1}H_{2}+h.c.)+(\tilep m_{14}^{2}H_{1}H_{4}+h.c.) \\ \nonumber
 &&+(\tilep m_{32}^{2}H_{3}H_{2}+h.c.)+(m_{34}^{2}H_{3}H_{4}+h.c.) \\ \nonumber
 &&+(\tilep m_{13}^{2}H_{1}^{\dagger}H_{3}+h.c.)
 +(\tilep m_{24}^{2}H_{2}^{\dagger}H_{4}+h.c.)+V_{D},
\end{eqnarray}
where $V_{D}$ represents the D-term and $m_{i}^{2}$ and $m_{ij}^{2}$ are
about $(100\sim 300{\rm~GeV})^{2}$.
Here $m_{ij}^{2}$ are complex and their phases cannot be removed by 
the redefinition of the Higgs fields and one phase is left.
In the MSSM or the NMSSM, all the phases of the Higgs mass parameters can be 
absorbed by field redefinitions and thus SCPV cannot occur without 
the appearance of a too light Higgs boson \cite{pomarol,romao,davies,haba}.
This is the reason for our model to be the minimal SCPV model.

These situations are the same as in Ref.\cite{masip-rasin} and 
as discussed there 
the potential Eq.(\ref{pottl}) has the vacuum with the following structure.
\begin{eqnarray}
 &&v_{1},\;v_{2}=O(v_{w}),\;\;\mbox{ real up to $\tilep^{2}$}, \\ \nonumber
 &&v_{3},\;v_{4}=O(\tilep v_{w}),\;\;\mbox{ arbitrary phases}.
\end{eqnarray}

Here we have taken the basis on which $m_{13}^{2}=m_{24}^{2}=0$ and 
$m_{12}$, $m_{32}$ and $m_{34}$ are real \cite{masip-rasin}.
On this basis, Yukawa couplings are
\begin{eqnarray}
 &&h^{u},\;h^{d},\;h^{e}\;\mbox{ : real up to $\tilep^{2}$}, \\ \nonumber
 &&h^{\prime u},\;h^{\prime d},\;h^{\prime e}\;\mbox{ : arbitrary phases}.
\end{eqnarray}
The hierarchical structure does not change by these field redefinitions.

Thus the CKM matrix is real up to $\tilep^{2}$.

\subsection{K physics}
The CP-violation parameter $\ep_{K}$ can be written in terms of 
the mass matrix element of the neutral K meson 
in the ${\rm K}^{0}$-$\bar{{\rm K}}^{0}$ basis 
$M_{12}$,
\begin{equation}
 |\ep_{K}|\simeq\frac{1}{2\sqrt{2}}\frac{{\rm Im}M_{12}}{{\rm Re}M_{12}}.
\end{equation}

\begin{figure}
 \leavevmode
 \epsfxsize=8cm
 \epsfysize=4cm
 \centerline{\epsfbox{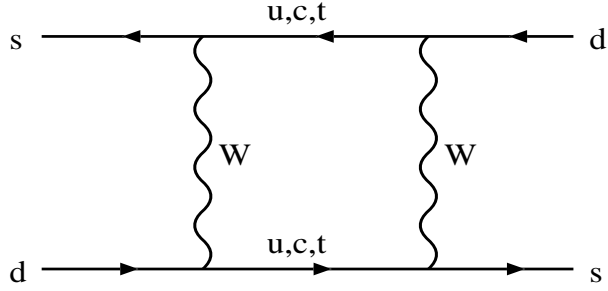}}
 \caption{The standard box diagram.}
 \label{boxdgm}
\end{figure}

\begin{figure}
 \leavevmode
 \epsfxsize=8cm
 \epsfysize=3cm
 \centerline{\epsfbox{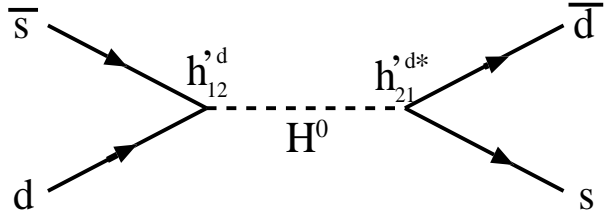}}
 \caption{The tree-level diagram with the neutral Higgs exchange.}
 \label{hgsex}
\end{figure}

The dominant contribution to ${\rm Re}M_{12}$ can come from 
the standard box diagram and the tree-level diagram with the neutral-Higgs
exchange depicted in Fig.~\ref{boxdgm} and Fig.~\ref{hgsex}.

The contribution of Fig.~\ref{boxdgm} is calculated as follows 
\cite{haber-nir}.

\begin{eqnarray}
 M_{12}^{\rm box}&=&\frac{G_{F}}{2\sqrt{2}}\frac{\alpha}
 {4\pi\sin^{2}\theta_{W}}
 (\cos\theta_{c}\sin\theta_{c})^{2}\frac{m_{c}^{2}}{m_{W}^{2}}
 \frac{\langle K^{0}|\bar{d}_{L}\gamma_{\mu}s_{L}\bar{d}_{L}\gamma^{\mu}s_{L}
 |\bar{K}^{0}\rangle}{m_{K}} \nonumber \\
 &\simeq& 10^{-13}\cdot\frac{\langle K^{0}|\bar{d}_{L}\gamma_{\mu}s_{L}
 \bar{d}_{L}\gamma^{\mu}s_{L}|\bar{K}^{0}\rangle}{m_{K}}\;({\rm GeV}^{-2}),
\end{eqnarray}
where $m_{c}$, $m_{W}$ and $m_{K}$ are the masses of the c quark, W boson
and the K meson respectively, and $G_{F}$ and $\alpha$ are the Fermi constant
and the fine structure constant.
$\theta_{W}$ and $\theta_{c}$ are the Weinberg angle and the Cabibbo angle
respectively.

On the other hand, the contribution of Fig.~\ref{hgsex} is calculated as
\begin{equation}
 M_{12}^{\rm tree}=\frac{h_{12}^{\prime d}h_{21}^{\prime d \ast}}
 {m_{H^{0}}^{2}}
 \frac{\langle K^{0}|\bar{d}_{L}s_{R}\bar{d}_{R}s_{L}|\bar{K}^{0}
 \rangle}{m_{K}},
\end{equation}
where $m_{H^{0}}$ is the neutral Higgs mass.

Then,
\begin{equation}
 \left|\frac{M_{12}^{\rm tree}}{M_{12}^{\rm box}}\right|\simeq 10^{13}\:
 ({\rm GeV}^{2})\:\frac{h_{12}^{\prime d}h_{21}^{\prime d \ast}}{m_{H^{0}}^{2}}
 \frac{\langle K^{0}|\bar{d}_{L}s_{R}\bar{d}_{R}s_{L}|\bar{K}^{0}
 \rangle}{\langle K^{0}|\bar{d}_{L}\gamma_{\mu}s_{L}\bar{d}_{L}\gamma^{\mu}
 s_{L}|\bar{K}^{0}\rangle}.
\end{equation}

According to Ref.\cite{haber-nir},
\begin{equation}
 \frac{\langle K^{0}|\bar{d}_{L}s_{R}\bar{d}_{R}s_{L}|\bar{K}^{0}
 \rangle}{\langle K^{0}|\bar{d}_{L}\gamma_{\mu}s_{L}\bar{d}_{L}\gamma^{\mu}
 s_{L}|\bar{K}^{0}\rangle}\simeq 7.6,
\end{equation}
and we obtain
\begin{equation}
 \left|\frac{M_{12}^{\rm tree}}{M_{12}^{\rm box}}\right|\simeq 
 \frac{7.6\times 10^{13}\:({\rm GeV}^{2})}{m_{H^{0}}^{2}}\cdot\frac{1}{40}
 \tilep\ep^{2}\cdot\frac{1}{40}\tilep\ep\simeq 150\,\tilep^{2}.
\end{equation}

Here we have assumed that $m_{H^{0}}=300$~GeV.
If we set $|v''|\simeq |v|\simeq 160$~TeV, $\tilep\simeq\ep\simeq 1/15$ and 
thus $|M_{12}^{\rm tree}/M_{12}^{\rm box}|=O(1)$.
In such a case, the value of $\ep_{K}$ becomes too large.
Then we will assume $|v''|\simeq |v|/2 \simeq 80$~TeV.
In this case, $\tilep\simeq 1/44$ and we obtain 
$|M_{12}^{\rm tree}/M_{12}^{\rm box}|=0.08$.
Therefore ${\rm Re}M_{12}$ is dominated by the box diagram.

On the contrary, the main contribution to ${\rm Im}M_{12}$ comes from 
the Fig.~\ref{hgsex} since the KM phase is greatly suppressed by the factor
$\tilep^{2}$ while the ``extra'' Yukawa couplings $h_{ij}^{\prime d}$ have 
arbitrary phases.

Hence we can estimate the absolute value of  $\ep_{K}$ at
\begin{equation}
 |\ep_{K}|\simeq\frac{1}{2\sqrt{2}}\left|\frac{M_{12}^{\rm tree}}
{M_{12}^{\rm box}}\right|\sim 10^{-2}.
\end{equation}

Taking into account the fact that there are $O(1)$ ambiguity in 
the hierarchical structure of the Yukawa matrices 
and the mixing among the neutral Higgs fields,
we can conclude that the above value of $\ep_{K}$ has an ambiguity 
about one order of magnitude.
As a result, we can estimate the value of $\ep_{K}$ at
\begin{equation}
 |\ep_{K}|\simeq 10^{-3}\mbox{ - }10^{-1},
\end{equation}
and this is consistent with the experimental value.

Next we will consider the value of $\ep_{K}'/\ep_{K}$.
$\ep_{K}'$ can be written as \cite{masip-rasin}
\begin{equation}
 |\ep_{K}'|\simeq\frac{1}{\sqrt{2}}\left|\frac{A_{2}}{A_{0}}\right|t_{0},\;\;
 \;\left(t_{0}\equiv\frac{{\rm Im}A_{0}}{{\rm Re}A_{2}}\right)
\end{equation}
where $A_{i}$ is the decay amplitude of a $K^{0}$ into two pions of isospin
$i$ and $|A_{2}/A_{0}|\simeq 1/22$ from the experiments.

The candidates of the main contributions to $t_{0}$ are the standard penguin
diagram, the penguin diagrams with chargino and stop and tree-level diagrams
with charged Higgs.
When the charged Higgs is relatively light, for example 
$m_{H^{+}}\simeq 150$~GeV, the third-type diagrams give the main contribution.
According to Ref.\cite{masip-rasin}, $\ep_{K}'/\ep_{K}$ can be estimated at
around $10^{-5}$ in the case of $m_{H^{+}}\sim 1$~TeV.
Then we can estimate $\ep_{K}'/\ep_{K}\sim 10^{-3}$ with 
$m_{H^{+}}\simeq 150$~GeV.
This value is also consistent with the experimental value.

\subsection{EDM and B physics}
For the same reason as discussed in \cite{masip-rasin}, the values of 
the electric dipole moments (EDM) of the neutron and the electron are 
below and close to the experimental upper bounds in our model.

Finally, we will comment that all the CP asymmetries in the B decays 
are small enough to be neglected because  
the KM phase and the phases of the ``standard'' Yukawa couplings are of order
$\tilep^{2}$, and 
the ``extra'' Yukawa couplings and the VEVs of the ``extra'' Higgs fields are
both suppressed by $\tilep$.

\section{Other phenomenological implications} \label{opi}
\subsection{Neutrino}
In our model, it does not seem that the see-saw mechanism can be applied
at first sight since the fundamental scale $M_{\ast}$ ($\simeq 1000$~TeV) is 
much lower than the scale required in the usual see-saw mechanism.
However, the volume factor suppression enables the neutrinos to have 
the desirable small masses.

Suppose that the right-handed neutrinos $\nu_{Ri}$ ($i=1,2,3$) are 
the bulk fields propagating into the extra dimension.
Then the neutrino Yukawa couplings $h_{ij}^{\nu}$ are suppressed by 
the volume factor: $1/2\pi M_{\ast}R\simeq 1/1600$.
Therefore it is possible to obtain the small neutrino masses of order eV
range even in the case that the Majorana mass scale of the right-handed 
neutrinos $M_{\nu_{R}}$ is around the low fundamental scale 
$M_{\ast}\simeq 1000$~TeV.
For example, if we assume $\nu_{Ri}$ not to have the $U(1)_{A}$ charges,
we can estimate the Dirac mass of the neutrino at
\begin{equation}
 m_{\nu}\simeq\frac{(\ep v_{w})^{2}}{M_{\nu_{R}}}\frac{1}{(2\pi M_{\ast}R)^{2}}
 \sim 10^{-1} \;\:{\rm (eV)},
\end{equation}
where $v_{w}=174$~GeV.
This is consistent with the value of the mass of $\nu_{\tau}$ estimated from
the neutrino-oscillation experiments.

Further, we can gain more suppression by assigning non-zero 
$U(1)_{A}$ charges to $\nu_{Ri}$.

\subsection{Strong CP problem}
We have a Nambu-Goldstone (NG) boson associated with the breaking of 
the $U(1)_{A}$ symmetry since $\bphi$ and $\bphi'$ have VEVs at $\mgut$ 
\footnote{To be exact, we have a {\it pseudo}-NG boson because $U(1)_{A}$ 
is anomalous.}.
This NG boson will behave like an axion, which can set the $\theta$-parameter
to be zero at low energy.

The axion field $\varphi$ couples to the quarks and leptons such as
\begin{equation}
 \frac{\lambda}{\mgut}\partial_{\mu}\varphi\bar{u}\gamma_{5}\gamma^{\mu}u,
\end{equation}
where $\lambda$ is a dimensionless $O(1)$ coupling and $u$ is the u-quark
field.
Since $\mgut\sim 10^{5}$~GeV here, this coupling seems too strong to
realize an invisible axion.

Then we will introduce a new extra dimension whose radius is denoted by $R_{2}$
and new fields $\psi$ and $\bar{\psi}$ that feel this new dimension.
To illustrate the situation, let us consider the toy model with the couplings 
such as
$\xi\varphi\bar{\psi}\psi$ and $\eta\varphi\bar{q}q$, where $\xi$ and $\eta$
are dimensionless couplings and $q$ denotes the quark field, and the bulk
fields $\psi$ and $\bar{\psi}$ couple to only $\varphi$.
Then the renormalization group equation (RGE) of $\xi$ is \cite{abel-king}
\begin{equation}
 4\pi\frac{{\rm d}\xi}{{\rm d}t_{2}}=2\pi t_{2}\xi^{3}.\;\;
 \left(t_{2}=t_{2}(\Lambda)\equiv\frac{1}{2\pi}\left(\frac{\Lambda}
 {R_{2}^{-1}}\right)\right)
\end{equation}

Solving this equation,
\begin{equation}
 \xi^{2}(\Lambda)=\frac{1}
 {1/\xi^{2}(R_{2}^{-1})-1/2\cdot(t_{2}^{2}-1/4\pi^{2})}.
\end{equation}

If we will set $t_{2}(\mgut)\sim 10^{5}$, i.e., $R_{2}^{-1}\sim 100$~MeV
and $\xi(\mgut)=O(1)$, we can obtain greatly suppressed coupling at low energy
$\xi(R_{2}^{-1})\sim 10^{-5}$.

On the other hand, the RGE of $\eta$ is 
\begin{equation}
 4\pi\frac{{\rm d}\eta}{{\rm d}t_{2}}=\eta\left(2\pi C_{\xi}t_{2}\xi^{2}
 -C_{e}\frac{g_{e}^{2}}{t_{2}}\right),
\end{equation}
where $C_{\xi}$ and $C_{e}$ are $O(1)$ constants and $g_{e}$ is 
the electro-magnetic gauge coupling.
We have neglected the terms involving small Yukawa couplings.

Solving this equation,
\begin{equation}
 \eta(\Lambda)\simeq C\cdot\frac{\xi^{C_{\xi}}(\Lambda)}
 {t_{2}^{C_{e}\cdot\alpha}},
\end{equation}
where $C$ is a constant determined by the initial condition.

In the case of $C, C_{\xi}=1$, we will obtain very small coupling 
$\eta(R_{2}^{-1})\simeq\xi(R_{2}^{-1})\sim 10^{-5}$.

This suppression of $\eta$ comes from the power-law running of 
the field renormalization factor of $\varphi$ \cite{sakamura}.
Thus the axion-quark coupling in our model $\lambda$ can receive 
the similar suppression and make the axion invisible at low energy.

\subsection{Higgs mass}
Our model becomes a supersymmetric 4-Higgs-doublet (4HD) standard model below 
the compactification scale $R^{-1}$, so the lightest Higgs boson must be
lighter than 130~GeV, which is the same upper bound as the MSSM case 
\cite{sakamura2}.
Note that there is no large contribution to the Higgs mass bound coming from 
the existence of the ``extra'' Higgs particles appearing at two-loop level 
discussed in Ref.\cite{sakamura2} in spite of the low cut-off scale 
$M_{\ast}$. 
This is because the ``extra'' Yukawa couplings are suppressed by $\tilep$ 
in our model.

\section{Conclusions} \label{concl}
Here we considered the minimal SUSY model in which the fermion mass hierarchy 
is realized by an abelian flavor symmetry $U(1)_{A}$ and the smallness
of the CP violation is controlled by the spontaneous CP violation 
in the context of the large extra dimensions.
In our model various physical scales such as the symmetry breaking scales of
$U(1)_{A}$, $Z_{2}$-parity that guarantees the natural flavor conservation
in the Higgs sector, CP symmetry and Peccei-Quinn symmetry that is 
identified with $U(1)_{A}$ are unified into the same scale
($\sim 100$~TeV), at which the grand unification of the gauge couplings
might be occur.
This is a quite attractive feature that should be possessed by the theory
beyond the standard model.
In models with the low fundamental scale $M_{\ast}$ like theories with
large extra dimensions, one must explain various phenomenological problems
only with scales that is below $M_{\ast}$.
Our model satisfies this requirement.

Our model becomes the 4HD SUSY standard model whose parameters are strongly
controlled by the high energy physics,
and we showed that it can realize the realistic size of the CP violation.

The small masses of the neutrinos can also be obtained by an assumption that 
the right-handed neutrinos are the bulk fields.

Furthermore, the strong CP problem can be solved by the presence of 
the NG boson associated with the $U(1)_{A}$ breaking, 
which is regarded as the axion and 
can be made invisible by introducing a new large extra dimension and 
new fields which couple to only the axion and feel this new dimension.

Finally, we comment about the hierarchy between the large VEVs of $\bphi$, 
$\bphi'$ and $\bphi''$ ($\sim 100$~TeV) and the small VEVs of the Higgs fields
($\sim 100$~GeV).
It is naturally understood by the fact that the physics on the boundary 
can be regarded as the fluctuation of the physics in the bulk 
and thus VEVs induced on the boundary are generically much smaller than VEVs 
in the bulk.

\vspace{1cm}
\mbox{}\linebreak
{\Large\bf Acknowledgements}

\mbox{}\linebreak
The author would like to thank N.Sakai and N.Maru for useful advice and 
careful checking of my idea and calculations.

\end{document}